%% file: main.tex
\documentclass[runningheads]{llncs}
\input{preamble}

\begin{document}
\title{Hashing Theorems Modulo Rewrites and Binder Permutation
Enables Equivalence Search over 200k Mathlib Theorems}
\titlerunning{Hashing Theorems Modulo Rewrites and Binder Permutation}
\author{Samuel Le{\ss}mann\inst{1}\textsuperscript{(\Letter)}\orcidlink{0009-0007-7845-3279} \and
  Simon Henniger\inst{2}\orcidlink{0009-0004-0574-2641} \and
  Gabriel Poesia\inst{3}\orcidlink{0009-0003-7058-8428} \and
  Nada Amin\inst{2}\orcidlink{0000-0002-0830-7248}}
\authorrunning{Le{\ss}mann, Henniger, Poesia, and Amin}
\institute{University of Cambridge, Cambridge, United Kingdom\\
  \email{sl2379@cam.ac.uk} \and
  Harvard University, Cambridge, MA, United States\\
  \email{shenniger@fas.harvard.edu}, \email{namin@seas.harvard.edu} \and
  University of Michigan, Ann Arbor, MI, United States\\
  \email{poesia@umich.edu}}
\maketitle
\SetWatermarkAngle{0}
\SetWatermarkText{\raisebox{11.7cm}{%
  \hspace{10.4cm}%
  \href{https://doi.org/10.5281/zenodo.23064352}{\includegraphics[width=20mm,keepaspectratio]{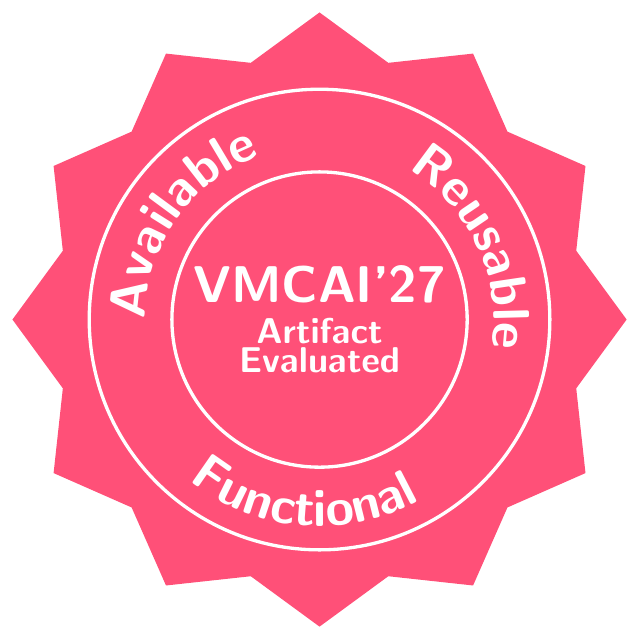}}%
}}
\begin{abstract}
  Like all code, Lean libraries accumulate technical debt and structural deficiencies over the course of their development. A particular concern is duplicate theorems: different parts of a big library can unwittingly contain and prove the same theorem, duplicating maintenance efforts. Tools can help detect redundancy and clean up. For instance, Mathlib's duplicate-declarations linter compares theorem types syntactically after sorting top-level binders and erasing binder information and universe levels. If detecting an equivalence requires unfolding, reorientation, or logical transformations, it escapes this comparison. In this work, we present a method to detect theorems which are equivalent with regard to argument order and currying/uncurrying of dependent types. We implement our method for Lean. By applying it to Mathlib, a library of more than 200k theorems, we demonstrate that it is computationally cheap and scalable. We discuss how our work may enhance efforts to tackle the maintenance burden of Lean libraries.

\end{abstract}

\section{Introduction}
\label{sec:intro}

As Lean grows in popularity, the amount of Lean code significantly increases. Consider Mathlib, which has grown to over 2.2 million lines of Lean. At that size, deprecation and
technical debt become central risks to continued library development~\cite{baanenGrowingMathlibMaintenance2025,doornMaintainingLibraryFormal2020}. Tooling for code maintenance becomes critical.
This issue is exacerbated with AI, which enables newer Lean projects to get even bigger even faster, e.g.
Anthropic's formalization of Fermat's Last Theorem (13 million lines)~\cite{anthropicFermatsLastTheoremLean2026,anthropicFormalizingFermatsLast2026} or
OpenAI's Lean certificates for finite-time blow-up of the
Navier--Stokes and Euler equations (641,000 lines)~\cite{openaiNavierStokesAndEuler2026,openaiNavierStokesMillennium2026}.

Duplicate theorems are of particular concern.
Two theorems in two different places that prove the same thing lead to confusion, bloat, and duplicated
community efforts.
However, given the expressivity of Lean's type system, automatically finding duplicates is a difficult problem. In general, proving arbitrary types equivalent is the same as proving arbitrary theorems and therefore cannot be decided algorithmically.

We propose a way for finding duplicates over a subset of type equivalences
which can be found by terminating syntactical rewriting.
To achieve this, we reuse some theoretical work on
type isomorphisms such as the complete axiomatizations of provable
isomorphism in typed
$\lambda$-calculi~\cite{bruceProvableIsomorphismsTypes1992,cosmoShortSurveyIsomorphisms2005,fioreRemarksIsomorphismsTyped2002},
their use in proof assistants~\cite{bartheTypeIsomorphismsProof2001},
and their extension to dependent
products~\cite{solovievIsomorphismDependentProducts2015}.
Instead of using them for search, however, we index entire Lean
libraries to find duplicates.

The relation we compute is deliberately narrower than logical equivalence. Two statements are identified only if a fixed set of isomorphism schemata and definitional unfoldings, applied at proposition positions, rewrites both to the same normal form up to renaming and reordering.
Because of this narrowness, the relation is cheap to decide over a
library of 200k Mathlib theorems while still finding relevant results.

\paragraph{Contributions.}
We make the following contributions:
\begin{itemize}
  \item We show that a terminating rewrite system with eight rules can find relevant duplicate theorems that
     a definitional equality check misses.
  \item Based on this insight, we implement a tool that finds duplicate
    statements in any Lean project and certifies their equivalence with a kernel-checked
    proof (\cref{sec:tool}).
    It normalizes statements, buckets the normal forms by an
    isomorphism-invariant signature, decides isomorphism within each bucket, and
    builds the certificates from generic transfer lemmas.
    Properties of its metatheory are formalized and proved on exactly
    the definitions the tool uses.
  \item We evaluate our tool on Lean core, Batteries and Mathlib libraries,
    on a total of 286{,}264 statements where we find 1{,}747
    redundant proofs (\cref{sec:eval}).
\end{itemize}

\section{Theoretical Foundations}
\label{sec:theory}

With our goal of working in a well-defined fragment of Lean's type theory,
we look for two properties.
First, we want to faithfully represent Lean propositions in the fragment and also be able to extract them back out.
Second, we want to have terminating normalization up to reordering and renaming of binders/hypotheses and be able to use efficient graph isomorphism heuristics to match them.
The connection between term equality with commutative structure and
graph isomorphism has been studied by
Basin~\cite{basinTermEqualityProblem1994}.

\subsection{Terms and Graphs}
\label{sec:terms}

Our term language \tm defined in \cref{fig:grammar-terms} is a syntactic
fragment of Lean covering terms,
types and propositions.
We use \tm not as its own type theory but as a syntactic fragment of
Lean propositions: it has no typing rules, no reduction, and every term is
obtained by encoding a Lean statement.
Hence, well-typedness is inherited from Lean and re-established
by the kernel when the readback is checked.
The design is similar to \cite{bruceProvableIsomorphismsTypes1992} and \cite{takimotoUnificationModuloIsomorphisms2025b}.
Since we only want to find duplicate theorems, only propositions are handled.
A $\forall$ whose body is a proposition, which also covers implication and negation, is opened by the rules, and a $\Pi$ whose codomain is not a
proposition is kept as an inert former, because isomorphisms between
types at arbitrary universes make the theory more complex than needed for our use.

\begin{figure}[ht]
  \input{parts/math/grammar_terms}
  \caption{
    Terms, types and propositions of the syntactic fragment \tm of
    Lean, with de~Bruijn
    indices for local binders and named pointers to graph nodes.
    Proof terms ($\star$, $\langle a, b \rangle_T$, $t.1$, $t.2$) are
    kept for the readback to Lean.
    The side conditions on $\forall$, $\Pi$, $\Sigma^{\wedge}$ and pair
    annotations are guaranteed by the encoder, not by the grammar.
    For readability, the examples name a bound variable $v_j$ after the depth $j$ of its binder counted from the outside instead of its index, and write $\mathsf{fv}(n)$ as $x_n$.
  }
  \label{fig:grammar-terms}
\end{figure}

We define a graph for representing our terms in~\cref{fig:grammar-graphs}
as a finite product of nodes and a node as a named variable with a type,
an arrow from a product of hypotheses to a product of conclusions,
a connective node with a product on each side, or a nameless equation node.
A product is a telescope in the sense of
de~Bruijn~\cite{debruijnTelescopicMappingsTyped1991}: later members
may refer to earlier ones by name, and the list order is a binder order.
A graph is well-formed when names are unique, labels are locally
closed, and every pointer targets a member that is in scope; the
initial one-node graph $x_0 : A$ of a statement without pointers is
well-formed, and every rewrite step preserves well-formedness.
\begin{figure}[tp]
  \input{parts/math/grammar_graphs}
  \caption{{Graphs \G are products of nodes, where a product is a
  telescope with named members and pointers between them.}}
  \label{fig:grammar-graphs}
\end{figure}

Two graphs are related by \iso iff a bijective renaming of names, applied to node names and pointers alike, followed by finitely many of the steps $\leftrightsquigarrow$ in \mbox{\cref{fig:iso}} turns one into the other.
A step exchanges two adjacent members of a product (\textsc{Swap}), the two sides of a connective node (\textsc{Flip}), or the two sides of an equation node (\textsc{FlipEq}).
Steps apply in every product, at the top level, in the domain or conclusion of an arrow, and on a side of a connective node, at any nesting depth, but never inside a label.
Hence an equation or connective that stays inside a label is compared only up to renaming.
\iso is the relation according to which normal forms are compared and decided.
On well-formed graphs, the rewrite system \R is proven to be confluent modulo this \iso in the single-source form stated in \mbox{\cref{sec:metatheory}}.
\begin{figure}[ht]
  \small
  \[
    \mathcal{C} \;\Coloneqq\; \square
    \;\mid\; N,\, \mathcal{C}
    \;\mid\; (\mathcal{C} \to c),\, g
    \;\mid\; (d \to \mathcal{C}),\, g
    \;\mid\; (\mathcal{C} \oplus c),\, g
    \;\mid\; (d \oplus \mathcal{C}),\, g
  \]
  \begin{mathpar}
    \inferrule*[right=Swap]
    { }
    {\mathcal{C}[\,N,\; N',\; g\,] \;\leftrightsquigarrow\;
    \mathcal{C}[\,N',\; N,\; g\,]}
    \and
    \inferrule*[right=Flip]
    { }
    {\mathcal{C}[\,(d \oplus c),\; g\,] \;\leftrightsquigarrow\;
    \mathcal{C}[\,(c \oplus d),\; g\,]}
    \and
    \inferrule*[right=FlipEq]
    { }
    {\mathcal{C}[\,a =_T b,\; g\,] \;\leftrightsquigarrow\;
    \mathcal{C}[\,b =_T a,\; g\,]}
    \and
    \inferrule*[right=Iso]
    {\varphi \text{ a bijection on names} \\
    \varphi(g) \;\leftrightsquigarrow^{*}\; g'}
    {g \mathrel{\text{\textbf{Iso}}} g'}
  \end{mathpar}
  \caption{
    The relation \iso and the product contexts $\mathcal{C}$ in which its steps and the rules of \R apply, where $N$ ranges over nodes, $g$, $d$ and $c$ over graphs, and $\oplus$ over $\vee$ and $\leftrightarrow$.
    $\varphi(g)$ renames every node name and every pointer of $g$ along $\varphi$.
    None of the three steps has a side condition.}
  \label{fig:iso}
\end{figure}

\subsection{The Rewrite System \R}
\label{sec:rewrite}
The eight rules of \R (\cref{fig:rules}) can be used at any member of any product, at the top level, in the domain or conclusion of an arrow, or on a side of a connective node, but never inside a label.
\textsc{Pi} and \textsc{Sigma} lift a binder out of a label into
the surrounding product while choosing fresh names.
The \textsc{ArrowMerge} rule
flattens nested arrows, representing currying.
\textsc{Unit}, \textsc{ArrowUnit} and \textsc{ArrowErase} remove $\top$, empty domains and arrows with an empty conclusion.
\textsc{Conn} and \textsc{Eq} turn $\vee$, $\leftrightarrow$ and $=$ into nodes for commutativity.
\textsc{Pi}, \textsc{Conn} and \textsc{Eq} apply only to a member
that no pointer in the whole graph references, i.e. a hypothesis that
is used somewhere else later is never split.
\begin{figure}[ht]
  \small
  \begin{mathpar}
    \inferrule*[right=Pi]
    {x, y \mathrel{\#} \mathcal{C}[\,x_i : \forall A.B,\; g\,] \\
    i \notin \mathsf{refs}(\mathcal{C}[\,x_i : \forall A.B,\; g\,])}
    {\mathcal{C}[\,x_i : \forall A.B,\; g\,] \;\to_{\mathbf{R}}\;
    \mathcal{C}[\,(x_x : A) \to (x_y : B\{x\}),\; g\,]}
    \and
    \inferrule*[right=Sigma]
    {x, y \mathrel{\#} \mathcal{C}[\,x_i : \Sigma^{\kappa}A.B,\; g\,]}
    {\mathcal{C}[\,x_i : \Sigma^{\kappa}A.B,\; g\,] \;\to_{\mathbf{R}}\;
      \mathcal{C}[\,x_x : A,\; x_y : B\{x\},\; g\,]\bigl[\langle
    \mathsf{fv}(x), \mathsf{fv}(y)\rangle_{\Sigma^{\kappa}A.B}\,/\,i\bigr]}
    \and
    \inferrule*[right=Conn]
    {x, y \mathrel{\#} \mathcal{C}[\,x_i : A \oplus B,\; g\,] \\
    i \notin \mathsf{refs}(\mathcal{C}[\,x_i : A \oplus B,\; g\,])}
    {\mathcal{C}[\,x_i : A \oplus B,\; g\,] \;\to_{\mathbf{R}}\;
    \mathcal{C}[\,(x_x : A) \oplus (x_y : B),\; g\,]}
    \and
    \inferrule*[right=Eq]
    {i \notin \mathsf{refs}(\mathcal{C}[\,x_i : a =_T b,\; g\,])}
    {\mathcal{C}[\,x_i : a =_T b,\; g\,] \;\to_{\mathbf{R}}\;
    \mathcal{C}[\,a =_T b,\; g\,]}
    \and
    \inferrule*[right=Unit]
    { }
    {\mathcal{C}[\,x_i : \top,\; g\,] \;\to_{\mathbf{R}}\;
    \mathcal{C}[\,g\,][\star/i]}
    \and
    \inferrule*[right=ArrowMerge]
    { }
    {\mathcal{C}[\,(d \to (d' \to c)),\; g\,]
      \;\to_{\mathbf{R}}\;
    \mathcal{C}[\,(d, d' \to c),\; g\,]}
    \and
    \inferrule*[right=ArrowUnit]
    { }
    {\mathcal{C}[\,(\cdot \to c),\; g\,]
    \;\to_{\mathbf{R}}\; \mathcal{C}[\,c,\; g\,]}
    \and
    \inferrule*[right=ArrowErase]
    { }
    {\mathcal{C}[\,(d \to \cdot),\; g\,]
    \;\to_{\mathbf{R}}\; \mathcal{C}[\,g\,]}
  \end{mathpar}
  \caption{
    The rewrite system \R on graphs where $\mathcal{C}[\cdot]$ is a
    product context (\cref{fig:iso}), $\#$ denotes freshness and $\mathsf{refs}$ the set
  of referenced names.}
  \label{fig:rules}
\end{figure}

\iso ignores the dependency order, so it does not preserve
well-formedness by itself.
However, what matters is that \R acts on \iso-classes: if $g$ is \iso
to $g'$ and $g$ steps to $h$, then $g'$ steps in one step to some
$h'$ \iso to $h$.
The tool only compares well-formed normal forms, so the member orders on both sides of an identified pair respect the dependencies.

Consider the two statements $T_1$ and $T_2$
\begin{align*}
  T_1 &\;=\; \forall m\,n : \mathbb{N}.\; 0 < n \to m\mid n \to m \le n \\
  T_2 &\;=\; \forall n\,m : \mathbb{N}.\; (\exists k.\, n = m \cdot k)
  \wedge 0 < n \to m \le n
\end{align*}
with $m\mid n$ unfolded to $\exists k.\, n = m \cdot k$.
Starting from the one-node graph $x_0 : T_1$, the rules of \R produce
\input{parts/math/normalization_T1}
and, from $x_0 : T_2$,
\input{parts/math/normalization_T2}

Both $N_{T_1}$ and $N_{T_2}$ are \R-normal: the remaining
variable nodes have atomic
types ($\mathbb{N}$, $0 < n$, $m \le n$) on which no rule applies.
They are related by permuting the members of the domain, and no renaming is needed because both traces choose the same names for corresponding nodes.
The tool generates the names instead, so its normal forms of $T_1$ and $T_2$ in \mbox{\cref{fig:pipeline}} are also related by a renaming.

\subsection{Metatheory Proved in Lean}
\label{sec:metatheory}

We formalize \tm, \R, \G and \iso and prove multiple important properties of them in Lean.
All theorems are formalized in a way such that the implementation
of the tool uses the same definitions we proved the properties about.
Our own definitions of \tm, \R and \iso are connected with the canonical
definitions from Mathlib and CSLib~\cite{barrettCSLibLeanComputer2026} for relations and rewrite systems.
They are proved in a separate folder and accepted by comparator~\cite{leanfroComparator2025}.
Termination of \R holds on all graphs, without a well-formedness
hypothesis, by a measure on spine formers and weights that every rule
decreases.
\iso is a simulation for \R in the sense of \cref{sec:rewrite}.
Well-formedness holds for the initial graph of every label without
pointers or loose bound variables and is preserved by every \R-step.
On well-formed graphs \R is confluent modulo \iso in the
single-source form.
If a well-formed $g$ reduces to $h_1$ and to $h_2$, then $h_1$ and
$h_2$ reduce to graphs related by \iso.
The abstract lemma is Newman's argument with \iso threaded through
under the simulation hypothesis, a special case of Huet's
confluence-modulo lemma.
Hence normal forms reachable from one well-formed graph are unique
up to \iso, and the normalizer, defined by well-founded recursion on
the termination theorem and without fuel, returns an \R-normal form
reached by \R-steps and, on well-formed graphs, one related by \iso
to every reachable normal form.
\iso is decidable with a reference decider proved total by structural
recursion, sound and complete on all graphs.
The key by which graphs are bucketed before comparison is invariant
under \iso.
All of these theorems depend on no axiom beyond the usual \texttt{propext},
\texttt{Classical.choice} and \texttt{Quot.sound}.
A one-page \texttt{Challenge.lean} file restates \G, \R, \iso, well-formedness
and the statements with only Mathlib and CSLib imported.

It is important to note that nothing proved in Lean
relates \R or \iso to the meaning of Lean propositions.
There is no semantics of the graphs and no adequacy or
completeness result for a theory of type isomorphisms.
The link to Lean is made per identified pair by the
equivalence certificates.

\section{Implementation}
\label{sec:tool}

Using our insights, we build a tool that scans Lean libraries for duplicates.
Importantly we use the same definitions for implementing the tool as we
use to prove the properties of \tm, \R, \G and \iso.
\cref{fig:pipeline} shows our tool's pipeline on the example from above.

\input{parts/figures/fig_pipeline_colour}

The tool scans all Lean statements which are propositions and
encodes them as a \tm term.
To find more duplicates, like in our example with $T_1$ and $T_2$, we need to unfold
$m \mid n$ to $\exists k.\, n = m \cdot k$.
We need to apply just the right amount of unfolding which is notoriously hard.
If we unfold too much, the term explodes in size.
Hence we only unfold definitions that are annotated as
\texttt{reducible} or \texttt{instanceReducible} (e.g.
definitions like $m \mid n$).

After encoding a statement as a \tm term and building the initial graph in \G with it, the rewrite system
\R is used to normalize it.
We previously proved that this always terminates.

In the default whole-library mode, the driver uses the proved-\iso-invariant signature to form buckets and the proved \iso decision procedure to compare members with class representatives.
The decision procedure first checks a witness proposed by the fast matcher and falls back to the complete reference decider when this attempt does not establish an isomorphism.
 Although there is no known polynomial-time algorithm for deciding
graph isomorphism, we can nicely use heuristic color refinement from
the one-dimensional Weisfeiler--Leman
algorithm~\cite{weisfeilerReductionGraphCanonical1968}.
Every node of a normal form starts with a color computed from its
label with all pointers blanked and from its position in the
graph.
A refinement round replaces the color of a node by a hash of its
old color, the colors of the nodes it points to, and, for every
incoming pointer, the color of the referrer together with the
position of the occurrence.
The rounds stop when no color class splits any more.
The stable colors become node hashes, the two sides of an
equation are combined commutatively, and the hashes of the members
of a telescope are summed, so the resulting signature does not depend
on the order of the members.
Inside each bucket, we compute the isomorphism with the node
hashes as a guide.
Equal signatures and pairwise distinct member hashes make the
search for a witness a sequence of forced moves without
backtracking.
On our data, the heuristic works in every tested case.

For each member $B$ of a bucket with representative $A$, the
tool builds a term of type $A \leftrightarrow B$ to certify that
both statements are indeed equivalent.
The certificate consists of a chain of generic equivalence
lemmas over arbitrary propositions and sorts, \texttt{Iff.refl},
\texttt{Iff.trans} and \texttt{Iff.symm}, the logical constants,
and readbacks of the intermediate graphs.
It references neither of the two theorems.
Since both statements are theorems, $A \leftrightarrow B$ is trivially provable from their proofs.
The certificate avoids this by using neither proof, so it would type-check even if $A$ and $B$ were unproved, and it witnesses their structural equivalence rather than the fact that both hold.

The relation is strictly finer than logical equivalence and is
generated by the schemata above and by definitional unfolding, nothing else.
Not identified are, for example, distributivity of $\forall$ and
$\to$ over $\wedge$, idempotence, De~Morgan laws, $\leftrightarrow$
as two implications, the one-point rule, unfolding of plain
definitions such as \texttt{Function.Injective}, permutation of
universe parameters, and symmetry of \texttt{HEq}.
A negative verdict therefore means only that \R and \iso do not
identify the two encodings, not that the statements are inequivalent.

\section{Evaluation}
\label{sec:eval}
To evaluate our tool, we ran it on a corpus of 286,264 theorem
statements from Lean core, Batteries, and Mathlib, including
deprecated aliases and generated API variants\footnote{Specifically, source revision \texttt{23c81b5} on Lean v4.35.0-rc2, Batteries \texttt{ed9b316}, and Mathlib \texttt{0653561}.
We used a general-purpose virtual private server with eight vCPUs and 32 GB memory.}.
On our hardware, the tool run took about half an hour, most of which was spent on certificate construction and kernel checking. The linter comparison and report generation took another 15 minutes.

\subsection{Duplicate Classes and Certificates}

\cref{tab:run-details} summarizes the run.
Of the 286{,}264 extracted statements, 286{,}224 were encoded and normalized within the configured size budgets and 40 were never compared.
This measures processing coverage of the selected corpus, independently of the equivalences omitted by the fixed schemata.
The normalized statements fall into 7{,}525 duplicate classes with at least two members, 15{,}834 members in total.
Pairing every non-representative member with its class representative gives 8{,}309 pairs, each with its own kernel-checked certificate.
Of these 8{,}309 members, 6{,}293 are definitionally equal to their representative according to \texttt{Meta.isDefEq}, so 76\,\% of the certificates are for pairs that Lean already accepts as definitionally equal; 2{,}016 members are not.

\input{parts/tables/tab_run}

Not every pair is a redundant proof.
We distinguish statement pairs certified by the kernel from the analysis scripts' labels about proof independence and likely maintenance value. A relevant finding is a redundant independent proof, counted within a class of statements equal modulo the isomorphisms.
If the directed graph of references between class members has $k$ proof roots, the class contributes $k-1$ relevant findings, where a root has a proof of its own, reaches no other member, is not generated from another member, and is not deprecated.
A proof generated from another member by \texttt{alias} or \texttt{to\_additive}, or generated together with it by the same command, is excluded.
A handwritten proof remains eligible when its statement is also generated by a macro or attribute from a third declaration.

Counted per class this gives 1{,}747 redundant proofs in 1{,}687
  classes, 965 of them for definitionally equal statements and 782 for
  statements equal only through the isomorphisms.

The automated classification labels 874 of the 1{,}747 redundant proofs as deliberate API twins, leaving 873 candidates in 844 classes outside that category.
These labels prioritize inspection and do not establish that the remaining declarations were introduced unintentionally or should be removed.

\subsection{Example Findings}

The pair below is an orientation twin of two theorems in the same file equal after flipping the two sides of $=$.
Both are there on purpose but they get flagged because the second
statement does not use the proof of the first.
Its independent proof could be replaced by \texttt{map\_zipIdx.symm}
to remove duplication and signify the connection between both theorems.

\begingroup\leanlistingstyle
\input{parts/listings/finding_list_zipidx}
\endgroup

A pair spanning two files of Mathlib, identified after
  re-associating the existential and the conjunction and reordering the
hypotheses, is the following:

\begingroup\leanlistingstyle
\input{parts/listings/finding_tensorproduct}
\endgroup
Both statements express that two tensors with equal images in the ambient tensor product already have equal images in some larger finitely generated submodule.
Their proofs use different presentations of the same direct-limit argument, which makes the pair a candidate for proof reuse across files.
This pair illustrates why comparison must account for dependencies as well as binder order, since the existential inclusion proof refers to the chosen submodule.
The certificate supplies a structural connection between the statements, while deciding whether to share their proofs requires a separate review of the library's API and import dependencies.

The classification labels are assigned automatically using rules based on the source text, proof terms, and Lean environment.
These labels have not been checked by hand, so the maintenance value of the reported candidates remains to be assessed.
Informal inspection of the findings suggests that a substantial number of the 873 candidates not labeled as deliberate API twins offer opportunities to improve code quality through proof reuse and clearer connections between declarations.
Inspection also reveals variation in binder order, the use of auto-implicit binders (outside Mathlib, which does not use them) and separately declared variables.
The findings also expose inconsistencies in theorem naming conventions.

\subsection{Comparison with Mathlib's Duplicate-Declarations Linter}

Mathlib's \texttt{DuplicateDecls} is a syntactic tool that groups theorem types by equal keys after erasing universe levels and binder information and sorting top-level binders.
It compares the resulting terms up to binder names, without unfolding definitions, reorienting connectives, or applying logical transformations.
We applied Mathlib's \texttt{DuplicateDecls} comparison and filtering rules to the same environment.
Of the 1{,}742 independent pairs reported by our analysis, the linter finds 481 (27.6\,\%), excludes 241 (13.8\,\%), and misses 1{,}020 (58.6\,\%).
This comparison counts representative--member pairs, whereas the 1{,}747 redundant proofs above are counted from proof roots per class.

For an excluded pair, at least one declaration is filtered out, with 160 pairs excluded only by the metaprogramming-name filter, 78 only by the alias filter, and 3 by both.
The metaprogramming filter skips namespaces such as \texttt{Lean.Grind}, \texttt{Lean.Omega}, and \texttt{Mathlib.Tactic}.
The alias filter skips proofs consisting of a single constant applied to the theorem's hypotheses, which also removes genuine one-line re-proofs.
For 118 of these 241 pairs, the comparison keys would agree if the filters were disabled.
This selection policy sensibly focuses reports on mathematical declarations and avoids routine aliases, so we distinguish deliberate exclusions from key misses and report overlap rather than overall recall.

Among the 1{,}020 pairs retained but missed by the linter, 527 are diagnosed as orientation changes and 316 are definitionally equal despite different keys.
The remaining 177 involve binder order, unfolding, currying, annotations, or other rewrites.
These diagnoses depend on the order of key relaxations rather than uniquely identifying each cause.

Conversely, our tool identifies 646 of the linter's 782 theorem groups entirely within single duplicate classes.
Of the 809 pairs of theorems within these groups, 142 are not identified by our tool, of which 119 fall outside our corpus and 23 differ in universe levels, which the linter erases and our tool preserves.

\section{Related Work}
\label{sec:related}

\paragraph{Type isomorphisms and theorem search.}
Using types as library search keys modulo isomorphisms dates
back to Rittri~\cite{rittriUsingTypesSearch1989}.
SearchIsos applies this approach to dependent types in
Rocq~\cite{delahayeInformationRetrievalCoq2000}, while Loogle
supports pattern-based theorem search in
Lean~\cite{joachimbreitnerLoogle2023}.
Takimoto et al.\ compare the search capabilities of these tools
and those of Rocq, Idris and Lambdapi, and propose a
semi-algorithm for unification modulo isomorphisms of dependent
products and sums~\cite{takimotoUnificationModuloIsomorphisms2025b}.
Their approach combines argument reordering and
currying/uncurrying with metavariable instantiation, allowing
queries to match more general statements or statements with
additional premises.
Our work addresses a different task: grouping existing theorem
statements across an entire library under a fixed structural
equivalence, without metavariable-based generalization.
We adapt their rewrite schemata to an annotated graph
representation, adding the arrow-to-unit law and explicit
symmetry for propositional connectives and equality.

\paragraph{Duplicate detection in Lean.}
Mathlib's \texttt{DuplicateDecls} is the closest existing tool
in purpose.
It compares declaration types syntactically after sorting top-level binders
and erasing binder information and universe levels, and
filters aliases to avoid reporting intentional duplicates.
Our approach extends comparison to structural transformations
within statements, including currying/uncurrying, decomposition
of conjunctions and existentials, and connective symmetry.
Dependency graphs retain references between binders while
allowing normal forms to be compared modulo renaming and
permutation.
We also selectively unfold definitions at proposition
positions.
Each reported match is accompanied by a kernel-checked
equivalence proof constructed from the rewrite and
isomorphism witnesses, without referencing either theorem.

\paragraph{Foundations and scope.}
Our approach builds on work on provable type isomorphisms in
typed $\lambda$-calculi~\cite{bruceProvableIsomorphismsTypes1992}
and their extension to dependent
products~\cite{solovievIsomorphismDependentProducts2015}.
The use of rewriting modulo an equivalence follows the
classical rewriting-modulo
setting~\cite{petersonCompleteSetsReductions1981}.
Our mechanized results establish termination and confluence
modulo \iso for the chosen graph rules, together with a
decision procedure for \iso.
These results concern the graph representation: they do not
establish completeness for a general theory of type
isomorphisms or logical equivalence.
The connection to Lean propositions is checked separately for
each reported pair through its equivalence certificate.

\section{Conclusion}
We presented our tool for detecting duplicate theorem statements through structural rewrites and dependency-preserving binder permutations.
Its rewrite system and graph-isomorphism decision procedure have a machine-checked metatheory, and each reported pair comes with a kernel-checked equivalence certificate.
The evaluation on 286{,}264 theorem statements demonstrates library-scale detection and exposes 873 maintenance candidates that our labels suggest are worth inspecting for opportunities to improve code quality.
These findings offer concrete opportunities to review proof reuse and make connections between declarations explicit.
The certificates establish statement equivalence, while deciding which proofs to share or declarations to retain requires careful human review.

As a next step one could integrate the certified duplicate reports into
library maintenance workflows, like for example Mathlib's technical debt tracking,
and evaluate the resulting proof refactorings with maintainers in more detail.
Another direction to extend the project would be to refine the supported schemata and
certificate construction beyond propositions to general isomorphisms of types.
Further, it would be interesting to find not just equivalences but also specializations or implications.
For example when an equivalence $A \leftrightarrow B$ is already proved, a proof of $A \to B$ might also be redundant.

\begin{credits}
  \subsubsection{Data Availability Statement.}
  The artifact for this paper, a Docker image of the tool with a full reproduction of the evaluation, is available on Zenodo~\cite{lessmannLeanPisoTripArtifact2026}.
  \subsubsection{\ackname}
  This research was partially supported by a Coefficient Giving grant. Simon Henniger was supported by the Harvard John A. Paulson School of Engineering and Applied Sciences (SEAS) Prize Fellowship and the German Academic Fellowship Organization, funded by the German Federal Ministry for Economic Affairs and Energy.\par
\end{credits}

\bibliographystyle{splncs04}
\bibliography{DepTypeIsoTeX}

\end{document}

%% file: preamble.tex
\expandafter\let\csname vec \endcsname\relax \let\vec\undefined
\DeclareMathAccent{\vec}{\mathord}{letters}{"7E}

\usepackage[T1]{fontenc}
\usepackage{newtxtext}
\usepackage[varvw]{newtxmath}       

\usepackage{graphicx}
\usepackage{mathpartir}     

\usepackage{fvextra}
\input{parts/listings/pygments_style}
\def\PYGZhy{\mbox{-}}       
\begingroup\catcode`\'=\active\gdef\PYGZsq{'}\endgroup  
\newcommand{\leanlistingstyle}{%
  \fvset{fontsize=\footnotesize,breaklines,xleftmargin=1em}%
  \VerbatimPygments{\PYG}{\PYG}}
\input{parts/leanunicode}

\usepackage{xcolor}
\usepackage[hidelinks]{hyperref}
\hypersetup{%
  pdftitle={Hashing Theorems Modulo Rewrites and Binder Permutation Enables Equivalence Search over 200k Mathlib Theorems},%
  pdfauthor={Samuel Le{\ss}mann, Simon Henniger, Gabriel Poesia, Nada Amin}}
\usepackage{orcidlink}
\usepackage{marvosym}       
\usepackage[firstpage]{draftwatermark}      

\usepackage{booktabs}
\usepackage{tikz}
\usetikzlibrary{matrix,fit,positioning,backgrounds,arrows.meta,calc}

\usepackage{cleveref}
\crefname{figure}{Fig.}{Figs.}
\Crefname{figure}{Figure}{Figures}
\crefname{section}{Sect.}{Sects.}
\Crefname{section}{Section}{Sections}
\crefname{subsection}{Sect.}{Sects.}
\Crefname{subsection}{Section}{Sections}
\crefname{subsubsection}{Sect.}{Sects.}
\Crefname{subsubsection}{Section}{Sections}
\crefname{table}{Table}{Tables}
\Crefname{table}{Table}{Tables}
\crefname{equation}{Eq.}{Eqs.}
\Crefname{equation}{Equation}{Equations}
\crefname{theorem}{Theorem}{Theorems}
\crefname{lemma}{Lemma}{Lemmas}
\crefname{definition}{Definition}{Definitions}
\crefname{proposition}{Proposition}{Propositions}
\crefname{corollary}{Corollary}{Corollaries}
\crefname{example}{Example}{Examples}

\usepackage{xspace}
\newcommand{\tm}{\(\mathbf{\text{\textbf{Tm}}}\)\xspace}
\newcommand{\R}{\(\mathbf{\text{\textbf{R}}}\)\xspace}
\newcommand{\G}{\(\mathbf{\text{\textbf{G}}}\)\xspace}
\newcommand{\iso}{\(\mathbf{\text{\textbf{Iso}}}\)\xspace}

%% file: parts/leanunicode.tex
\usepackage{stmaryrd}                                        
\DeclareFontFamily{U}{stmry}{}
\DeclareFontShape{U}{stmry}{m}{n}{%
  <-5.5> stmary5 <5.5-6.5> stmary6 <6.5-7.5> stmary7 <7.5-8.5> stmary8
  <8.5-9.5> stmary9 <9.5-> stmary10}{}
\DeclareFontShape{U}{stmry}{b}{n}{%
  <-5.5> stmary5 <5.5-6.5> stmary6 <6.5-7.5> stmary7 <7.5-8.5> stmary8
  <8.5-9.5> stmary9 <9.5-> stmary10}{}
\DeclareUnicodeCharacter{2200}{\ensuremath{\forall}}         
\DeclareUnicodeCharacter{2203}{\ensuremath{\exists}}         
\DeclareUnicodeCharacter{2227}{\ensuremath{\wedge}}          
\DeclareUnicodeCharacter{2228}{\ensuremath{\vee}}            
\DeclareUnicodeCharacter{00AC}{\ensuremath{\neg}}            
\DeclareUnicodeCharacter{22A4}{\ensuremath{\top}}            
\DeclareUnicodeCharacter{22A5}{\ensuremath{\bot}}            
\DeclareUnicodeCharacter{22A2}{\ensuremath{\vdash}}          
\DeclareUnicodeCharacter{2192}{\ensuremath{\rightarrow}}     
\DeclareUnicodeCharacter{2190}{\ensuremath{\leftarrow}}      
\DeclareUnicodeCharacter{2194}{\ensuremath{\leftrightarrow}} 
\DeclareUnicodeCharacter{21A6}{\ensuremath{\mapsto}}         
\DeclareUnicodeCharacter{2191}{\ensuremath{\uparrow}}        
\DeclareUnicodeCharacter{2193}{\ensuremath{\downarrow}}      
\DeclareUnicodeCharacter{21D2}{\ensuremath{\Rightarrow}}     
\DeclareUnicodeCharacter{21D4}{\ensuremath{\Leftrightarrow}} 
\DeclareUnicodeCharacter{2264}{\ensuremath{\leq}}            
\DeclareUnicodeCharacter{2265}{\ensuremath{\geq}}            
\DeclareUnicodeCharacter{2260}{\ensuremath{\neq}}            
\DeclareUnicodeCharacter{2243}{\ensuremath{\simeq}}          
\DeclareUnicodeCharacter{2245}{\ensuremath{\cong}}           
\DeclareUnicodeCharacter{2248}{\ensuremath{\approx}}         
\DeclareUnicodeCharacter{2261}{\ensuremath{\equiv}}          
\DeclareUnicodeCharacter{2208}{\ensuremath{\in}}             
\DeclareUnicodeCharacter{2209}{\ensuremath{\notin}}          
\DeclareUnicodeCharacter{2286}{\ensuremath{\subseteq}}       
\DeclareUnicodeCharacter{2282}{\ensuremath{\subset}}         
\DeclareUnicodeCharacter{2288}{\ensuremath{\nsubseteq}}      
\DeclareUnicodeCharacter{2223}{\ensuremath{\mid}}            
\DeclareUnicodeCharacter{2224}{\ensuremath{\nmid}}           
\DeclareUnicodeCharacter{2297}{\ensuremath{\otimes}}         
\DeclareUnicodeCharacter{2295}{\ensuremath{\oplus}}          
\DeclareUnicodeCharacter{00D7}{\ensuremath{\times}}          
\DeclareUnicodeCharacter{2218}{\ensuremath{\circ}}           
\DeclareUnicodeCharacter{2022}{\ensuremath{\bullet}}         
\DeclareUnicodeCharacter{00B7}{\ensuremath{\cdot}}           
\DeclareUnicodeCharacter{222A}{\ensuremath{\cup}}            
\DeclareUnicodeCharacter{2229}{\ensuremath{\cap}}            
\DeclareUnicodeCharacter{2211}{\ensuremath{\sum}}            
\DeclareUnicodeCharacter{220F}{\ensuremath{\prod}}           
\DeclareUnicodeCharacter{222B}{\ensuremath{\int}}            
\DeclareUnicodeCharacter{221E}{\ensuremath{\infty}}          
\DeclareUnicodeCharacter{2205}{\ensuremath{\emptyset}}       
\DeclareUnicodeCharacter{2207}{\ensuremath{\nabla}}          
\DeclareUnicodeCharacter{2202}{\ensuremath{\partial}}        
\DeclareUnicodeCharacter{221A}{\ensuremath{\surd}}           
\DeclareUnicodeCharacter{2016}{\ensuremath{\|}}              
\DeclareUnicodeCharacter{2A06}{\ensuremath{\bigsqcup}}       
\DeclareUnicodeCharacter{2A05}{\ensuremath{\bigsqcap}}       
\DeclareUnicodeCharacter{2294}{\ensuremath{\sqcup}}          
\DeclareUnicodeCharacter{2293}{\ensuremath{\sqcap}}          
\DeclareUnicodeCharacter{27E8}{\ensuremath{\langle}}         
\DeclareUnicodeCharacter{27E9}{\ensuremath{\rangle}}         
\DeclareUnicodeCharacter{27E6}{\ensuremath{\llbracket}}      
\DeclareUnicodeCharacter{27E7}{\ensuremath{\rrbracket}}      
\DeclareUnicodeCharacter{2983}{\ensuremath{\{\!\!\{}}        
\DeclareUnicodeCharacter{2984}{\ensuremath{\}\!\!\}}}        
\DeclareUnicodeCharacter{230A}{\ensuremath{\lfloor}}         
\DeclareUnicodeCharacter{230B}{\ensuremath{\rfloor}}         
\DeclareUnicodeCharacter{2308}{\ensuremath{\lceil}}          
\DeclareUnicodeCharacter{2309}{\ensuremath{\rceil}}          
\DeclareUnicodeCharacter{2115}{\ensuremath{\mathbb{N}}}      
\DeclareUnicodeCharacter{2124}{\ensuremath{\mathbb{Z}}}      
\DeclareUnicodeCharacter{211A}{\ensuremath{\mathbb{Q}}}      
\DeclareUnicodeCharacter{211D}{\ensuremath{\mathbb{R}}}      
\DeclareUnicodeCharacter{2102}{\ensuremath{\mathbb{C}}}      
\DeclareUnicodeCharacter{03B1}{\ensuremath{\alpha}}          
\DeclareUnicodeCharacter{03B2}{\ensuremath{\beta}}           
\DeclareUnicodeCharacter{03B3}{\ensuremath{\gamma}}          
\DeclareUnicodeCharacter{03B4}{\ensuremath{\delta}}          
\DeclareUnicodeCharacter{03B5}{\ensuremath{\varepsilon}}     
\DeclareUnicodeCharacter{03B6}{\ensuremath{\zeta}}           
\DeclareUnicodeCharacter{03B7}{\ensuremath{\eta}}            
\DeclareUnicodeCharacter{03B8}{\ensuremath{\theta}}          
\DeclareUnicodeCharacter{03B9}{\ensuremath{\iota}}           
\DeclareUnicodeCharacter{03BA}{\ensuremath{\kappa}}          
\DeclareUnicodeCharacter{03BB}{\ensuremath{\lambda}}         
\DeclareUnicodeCharacter{03BC}{\ensuremath{\mu}}             
\DeclareUnicodeCharacter{03BD}{\ensuremath{\nu}}             
\DeclareUnicodeCharacter{03BE}{\ensuremath{\xi}}             
\DeclareUnicodeCharacter{03C0}{\ensuremath{\pi}}             
\DeclareUnicodeCharacter{03C1}{\ensuremath{\rho}}            
\DeclareUnicodeCharacter{03C3}{\ensuremath{\sigma}}          
\DeclareUnicodeCharacter{03C4}{\ensuremath{\tau}}            
\DeclareUnicodeCharacter{03C6}{\ensuremath{\varphi}}         
\DeclareUnicodeCharacter{03C7}{\ensuremath{\chi}}            
\DeclareUnicodeCharacter{03C8}{\ensuremath{\psi}}            
\DeclareUnicodeCharacter{03C9}{\ensuremath{\omega}}          
\DeclareUnicodeCharacter{0393}{\ensuremath{\Gamma}}          
\DeclareUnicodeCharacter{0394}{\ensuremath{\Delta}}          
\DeclareUnicodeCharacter{0398}{\ensuremath{\Theta}}          
\DeclareUnicodeCharacter{039B}{\ensuremath{\Lambda}}         
\DeclareUnicodeCharacter{03A0}{\ensuremath{\Pi}}             
\DeclareUnicodeCharacter{03A3}{\ensuremath{\Sigma}}          
\DeclareUnicodeCharacter{03A6}{\ensuremath{\Phi}}            
\DeclareUnicodeCharacter{03A8}{\ensuremath{\Psi}}            
\DeclareUnicodeCharacter{03A9}{\ensuremath{\Omega}}          
\DeclareUnicodeCharacter{207B}{\ensuremath{^{-}}}            
\DeclareUnicodeCharacter{2070}{\ensuremath{^{0}}}            
\DeclareUnicodeCharacter{00B9}{\ensuremath{^{1}}}            
\DeclareUnicodeCharacter{00B2}{\ensuremath{^{2}}}            
\DeclareUnicodeCharacter{00B3}{\ensuremath{^{3}}}            
\DeclareUnicodeCharacter{207F}{\ensuremath{^{n}}}            
\DeclareUnicodeCharacter{2080}{\ensuremath{_{0}}}            
\DeclareUnicodeCharacter{2081}{\ensuremath{_{1}}}            
\DeclareUnicodeCharacter{2082}{\ensuremath{_{2}}}            
\DeclareUnicodeCharacter{2083}{\ensuremath{_{3}}}            
\DeclareUnicodeCharacter{2084}{\ensuremath{_{4}}}            
\DeclareUnicodeCharacter{1D62}{\ensuremath{_{i}}}            
\DeclareUnicodeCharacter{2C7C}{\ensuremath{_{j}}}            
\DeclareUnicodeCharacter{2099}{\ensuremath{_{n}}}            
\DeclareUnicodeCharacter{2039}{\ensuremath{\langle}}         
\DeclareUnicodeCharacter{203A}{\ensuremath{\rangle}}         
\DeclareUnicodeCharacter{2026}{\ensuremath{\ldots}}          
\DeclareUnicodeCharacter{2032}{\ensuremath{'}}               
\DeclareUnicodeCharacter{2019}{'}                            
\DeclareUnicodeCharacter{00B0}{\ensuremath{^{\circ}}}        
\DeclareUnicodeCharacter{2118}{\ensuremath{\wp}}             

%% file: parts/math/grammar_terms.tex
\[
  \begin{array}{r@{\;}c@{\;}l@{\quad}l}
    \multicolumn{3}{l}{\text{term language \tm}} & \text{Lean reading}\\[1pt]
    \hline\\[-7pt]
    \kappa & \Coloneqq & \exists \mid \wedge
    & \text{the two readings of a }\Sigma\\
    \oplus & \Coloneqq & \vee \mid \leftrightarrow
    & \text{commutative binary connectives}\\[4pt]
    t, a, b, A, B, T & \Coloneqq & \underline{i}
    & \text{bound variable, de~Bruijn index } i\\
    & \mid & \mathsf{fv}(n)
    & \text{pointer to the variable node named } n\\
    & \mid & \mathsf{atom}_k
    & \text{entry } k \text{ of the atom table (closed, opaque)}\\
    & \mid & t\; t'
    & \text{application}\\
    & \mid & \lambda A.\, b
    & \mathtt{fun}\ x : A \Rightarrow b\\
    & \mid & \forall A.\, B
    & \forall\, x : A,\ B \ \text{ with } B : \mathsf{Prop}\ \text{ (also } A \to B,\ \neg A)\\
    & \mid & \Pi A.\, B
    & (x : A) \to B \ \text{ with } B \text{ not a proposition}\\
    & \mid & \Sigma^{\kappa} A.\, B
    & \exists\, x : A,\ B \ \text{ if } \kappa = \exists;\quad A \wedge B \ \text{ if } \kappa = \wedge\\
    & \mid & A \oplus B
    & A \vee B,\ A \leftrightarrow B\\
    & \mid & a =_T b
    & \mathtt{@Eq}\ T\ a\ b\\
    & \mid & \top
    & \mathtt{True}\\
    & \mid & \star
    & \mathtt{True.intro}\\
    & \mid & \langle a, b \rangle_T
    & \mathtt{And.intro}\ a\ b,\ \mathtt{Exists.intro}\ a\ b\ \text{ at the } \Sigma\text{-type } T\\
    & \mid & t.1 \mid t.2
    & \text{projections of } t : A \wedge B
  \end{array}
\]

%% file: parts/math/grammar_graphs.tex
\[
  \begin{array}{r@{\;}c@{\;}l@{\qquad}l}
    N & \Coloneqq & x_n : A
    & \text{variable node: name } n \text{ of type } A\\
    & \mid & \Delta \longrightarrow C
    & \text{arrow: hypotheses } \Delta \text{ entail conclusions } C\\
    & \mid & \Delta \oplus C
    & \text{connective on graphs, } \oplus \in \{\vee, \leftrightarrow\}\\
    & \mid & a =_T b
    & \text{equation at type } T\\[4pt]
    \Delta, C & \Coloneqq & N_1, \ldots, N_m
    & \text{graph: a finite product of nodes } (m \ge 0)
  \end{array}
\]

%% file: parts/math/normalization_T1.tex
\[
  \resizebox{\textwidth}{!}{$\displaystyle
  \begin{array}{r@{\;}l@{\qquad}l}
    & x_0 : T_1 \\
    \to_{\mathbf{R}} & (m : \mathbb{N}) \longrightarrow (h_1 :
    \forall n : \mathbb{N}.\, 0 < n \to \cdots) & \textsc{Pi} \\
    \to_{\mathbf{R}} & (m : \mathbb{N}) \longrightarrow ((n :
    \mathbb{N}) \longrightarrow (h_2 : 0 < n \to \cdots)) &
    \textsc{Pi}\ \text{(inside the conclusion)} \\
    \to_{\mathbf{R}} & (m : \mathbb{N},\, n : \mathbb{N})
    \longrightarrow (h_2 : 0 < n \to (\exists k.\, n = m\cdot k) \to
    m \le n) & \textsc{ArrowMerge} \\
    \to_{\mathbf{R}}^{4} & (m : \mathbb{N},\, n : \mathbb{N},\, p : 0
    < n,\, q : \exists k.\, n = m \cdot k) \longrightarrow (r : m \le
    n) & \textsc{Pi} \times 2, \textsc{ArrowMerge} \times 2 \\
    \to_{\mathbf{R}} & (m : \mathbb{N},\, n : \mathbb{N},\, p : 0 <
    n,\, k : \mathbb{N},\, e : n = m \cdot k) \longrightarrow (r : m
    \le n) & \textsc{Sigma}^{\exists} \\
    \to_{\mathbf{R}} & (m : \mathbb{N},\, n : \mathbb{N},\, p : 0 <
    n,\, k : \mathbb{N},\, n =_{\mathbb{N}} m \cdot k)
    \longrightarrow (r : m \le n) & \textsc{Eq} \;=:\; N_{T_1}
  \end{array}$}
\]

%% file: parts/math/normalization_T2.tex
\[
  \resizebox{\textwidth}{!}{$\displaystyle
  \begin{array}{r@{\;}l@{\qquad}l}
    & x_0 : T_2 \\
    \to_{\mathbf{R}}^{5} & (n : \mathbb{N},\, m : \mathbb{N},\, h :
    (\exists k.\, n = m \cdot k) \wedge 0 < n) \longrightarrow (r : m
    \le n) & \textsc{Pi} \times 3, \textsc{ArrowMerge} \times 2 \\
    \to_{\mathbf{R}} & (n : \mathbb{N},\, m : \mathbb{N},\, q :
    \exists k.\, n = m \cdot k,\, p : 0 < n) \longrightarrow (r : m
    \le n) & \textsc{Sigma}^{\wedge} \\
    \to_{\mathbf{R}} & (n : \mathbb{N},\, m : \mathbb{N},\, k :
    \mathbb{N},\, e : n = m \cdot k,\, p : 0 < n) \longrightarrow (r
    : m \le n) & \textsc{Sigma}^{\exists} \\
    \to_{\mathbf{R}} & (n : \mathbb{N},\, m : \mathbb{N},\, k :
    \mathbb{N},\, n =_{\mathbb{N}} m \cdot k,\, p : 0 < n)
    \longrightarrow (r : m \le n) & \textsc{Eq} \;=:\; N_{T_2}
  \end{array}$}
\]

%% file: parts/figures/fig_pipeline_colour.tex
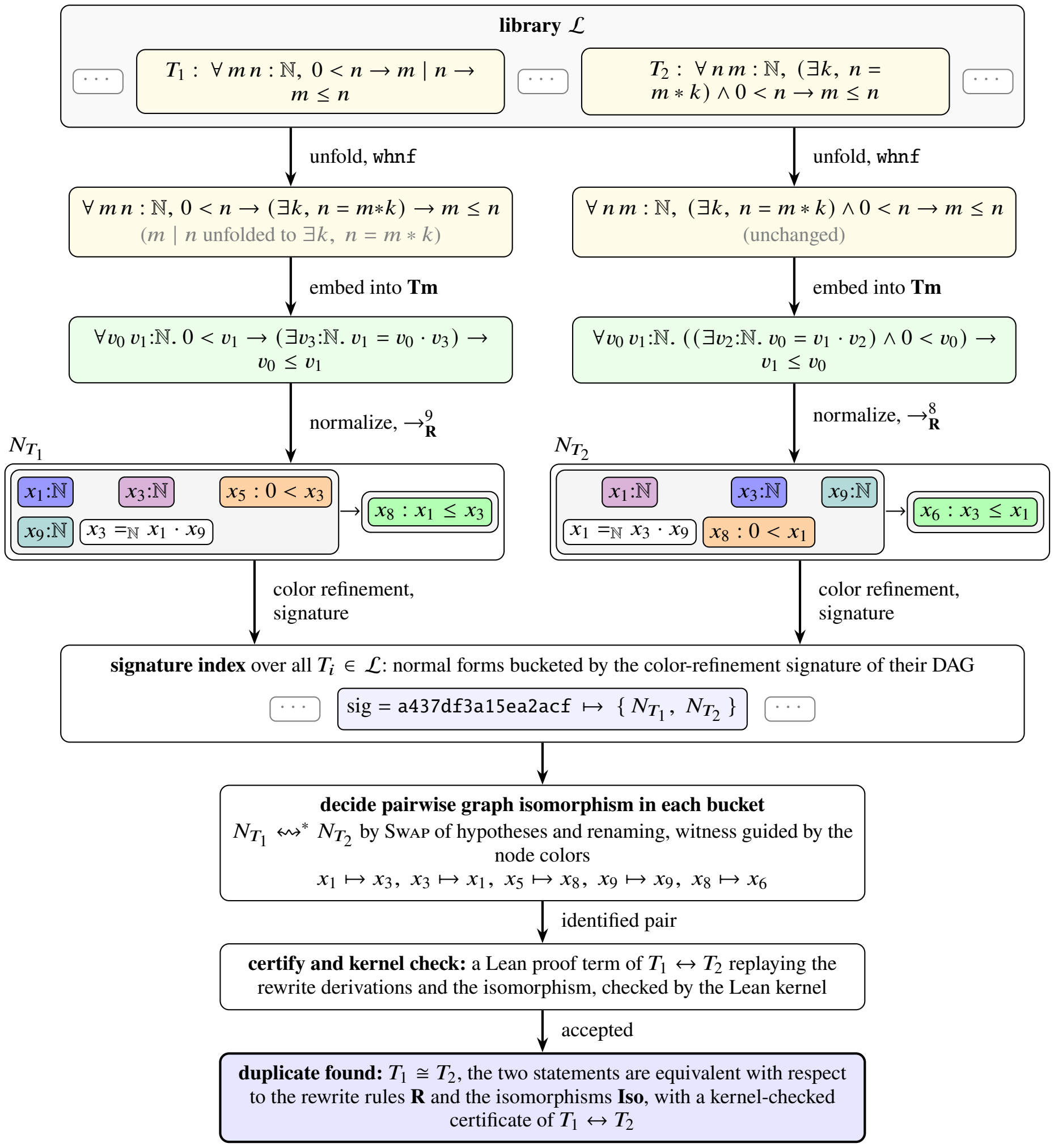
\begin{figure}[tp]
  \centering
  \newsavebox{\pipelinebox}%
  \begin{lrbox}{\pipelinebox}%
  \begin{tikzpicture}[
      font=\scriptsize,
      box/.style={draw, rounded corners=3pt, inner sep=4pt, align=flush center,
                  text width=50mm, fill=white},
      lean/.style={box, fill=yellow!10},
      tyb/.style={box, fill=green!8},
      wide/.style={box, text width=112mm},
      other/.style={draw, rounded corners=2pt, inner sep=3pt, fill=white,
                    text=gray, draw=gray},
      lane/.style={font=\scriptsize\bfseries},
      gnode/.style={draw, rounded corners=2pt, inner sep=2pt, fill=white},
      arr/.style={draw=none, inner xsep=2pt, inner ysep=0pt, fill=none},
      grp/.style={draw, rounded corners=3pt, inner sep=2pt, fill=gray!8},
      nf/.style={draw, rounded corners=3pt, inner sep=4pt, fill=none},
      dom/.style={matrix of math nodes, nodes={gnode, anchor=center},
                  column sep=2pt, row sep=3pt, inner sep=0pt},
      flow/.style={-{Stealth[length=5pt]}, thick},
      lbl/.style={midway, right=3pt, font=\scriptsize, align=left},
      final/.style={box, thick, fill=blue!10, text width=74mm},
      cP/.style={fill=blue!40},      
      cQ/.style={fill=violet!30},    
      cK/.style={fill=teal!30},      
      cLT/.style={fill=orange!35},   
      cLE/.style={fill=green!30},    
    ]
    \node[wide, fill=gray!5] (lib) at (0,0) {%
      \textbf{library} $\mathcal L$\\[3pt]
      \begin{tikzpicture}[baseline=(t1.base)]
        \node[other] (t0) {$\cdots$};
        \node[lean, text width=41mm, right=1.5mm of t0] (t1)
          {$T_1 :\ \forall\, m\,n : \mathbb N,\ 0<n \to m \mid n \to m\le n$};
        \node[other, right=1.5mm of t1] (tm) {$\cdots$};
        \node[lean, text width=41mm, right=1.5mm of tm] (t2)
          {$T_2 :\ \forall\, n\,m : \mathbb N,\ (\exists k,\ n = m*k)\wedge 0<n \to m\le n$};
        \node[other, right=1.5mm of t2] (tn) {$\cdots$};
      \end{tikzpicture}};

    \coordinate (cq) at ( 30mm, 0);
    \coordinate (cl) at (-30mm, 0);

    \node[lean] (q2) at ($(cq |- lib.south) + (0,-7mm)$) [anchor=north]
      {$\forall\, n\,m : \mathbb N,\ (\exists k,\ n = m*k)\wedge 0<n \to m\le n$\\[1pt]
       {\color{gray}(unchanged)}};
    \node[lean] (l2) at ($(cl |- lib.south) + (0,-7mm)$) [anchor=north]
      {$\forall\, m\,n : \mathbb N,\ 0<n \to (\exists k,\ n = m*k) \to m\le n$\\[1pt]
       {\color{gray}($m\mid n$ unfolded to $\exists k,\ n = m*k$)}};

    \node[tyb, below=7mm of q2] (q3)
      {$\forall v_0\,v_1{:}\mathbb N.\ ((\exists v_2{:}\mathbb N.\ v_0 = v_1\cdot v_2)\wedge 0<v_0)\to v_1\le v_0$};
    \node[tyb, below=7mm of l2] (l3)
      {$\forall v_0\,v_1{:}\mathbb N.\ 0<v_1 \to (\exists v_3{:}\mathbb N.\ v_1 = v_0\cdot v_3)\to v_0\le v_1$};

    \matrix (qd) [dom, below=11mm of q3.south, anchor=north east, xshift=10mm] {
      |[cQ]| x_1{:}\mathbb N & |[cP]| x_3{:}\mathbb N & |[cK]| x_9{:}\mathbb N \\
      x_1 =_{\mathbb N} x_3\cdot x_9 & |[cLT]| x_8 : 0<x_1 \\
    };
    \node[arr, right=0pt of qd] (qa) {$\rightarrow$};
    \node[gnode, cLE, right=0pt of qa] (qc) {$x_6 : x_3\le x_1$};

    \matrix (ld) [dom, below=11mm of l3.south, anchor=north east, xshift=5mm] {
      |[cP]| x_1{:}\mathbb N & |[cQ]| x_3{:}\mathbb N & |[cLT]| x_5 : 0<x_3 \\
      |[cK]| x_9{:}\mathbb N & x_3 =_{\mathbb N} x_1\cdot x_9 \\
    };
    \node[arr, right=0pt of ld] (la) {$\rightarrow$};
    \node[gnode, cLE, right=0pt of la] (lc) {$x_8 : x_1\le x_3$};

    \begin{scope}[on background layer]
      \node[nf, fit=(qd)(qa)(qc)] (qg) {};
      \node[nf, fit=(ld)(la)(lc)] (lg) {};
      \node[grp, fit=(qd)] (qgd) {};  \node[grp, fit=(qc)] (qgc) {};
      \node[grp, fit=(ld)] (lgd) {};  \node[grp, fit=(lc)] (lgc) {};
    \end{scope}
    \node[font=\scriptsize\itshape, anchor=south west, inner sep=1pt] at (qg.north west) {$N_{T_2}$};
    \node[font=\scriptsize\itshape, anchor=south west, inner sep=1pt] at (lg.north west) {$N_{T_1}$};

    \coordinate (mid) at ($(qg.south)!0.5!(lg.south)$);
    \node[wide] (idx) at ($(0,0 |- mid) + (0,-10mm)$) [anchor=north] {%
      \textbf{signature index} over all $T_i \in \mathcal L$: normal forms bucketed by the
      color-refinement signature of their DAG\\[3pt]
      \begin{tikzpicture}[baseline=(b1.base)]
        \node[other] (b0) {$\cdots$};
        \node[draw, rounded corners=2pt, inner sep=3pt, fill=blue!6, right=2mm of b0] (b1)
          {$\mathrm{sig} = \text{\ttfamily a437df3a15ea2acf}
            \ \mapsto\ \{\,N_{T_1},\ N_{T_2}\,\}$};
        \node[other, right=2mm of b1] (b3) {$\cdots$};
      \end{tikzpicture}};

    \node[box, text width=74mm, below=5mm of idx] (iso)
      {\textbf{decide pairwise graph isomorphism in each bucket}\\[1pt]
       $N_{T_1} \leftrightsquigarrow^{*} N_{T_2}$ by \textsc{Swap} of
       hypotheses and renaming, witness guided by the node colors\\
       $x_1\mapsto x_3,\ x_3\mapsto x_1,\ x_5\mapsto x_8,\ x_9\mapsto x_9,\ x_8\mapsto x_6$};
    \node[box, text width=74mm, below=5mm of iso] (cert)
      {\textbf{certify and kernel check:} a Lean proof term of
       $T_1 \leftrightarrow T_2$ replaying the rewrite derivations and the
       isomorphism, checked by the Lean kernel};
    \node[final, below=5mm of cert] (dup)
      {\textbf{duplicate found:} $T_1 \cong T_2$, the two statements are
       equivalent with respect to the rewrite rules \R and the isomorphisms
       \iso, with a kernel-checked certificate of $T_1 \leftrightarrow T_2$};

    \draw[flow] (cq |- lib.south) -- (q2.north) node[lbl] {unfold, \texttt{whnf}};
    \draw[flow] (cl |- lib.south) -- (l2.north) node[lbl] {unfold, \texttt{whnf}};
    \draw[flow] (q2) -- (q3) node[lbl] {embed into \tm};
    \draw[flow] (l2) -- (l3) node[lbl] {embed into \tm};
    \draw[flow] (q3.south) -- (q3.south |- qg.north)
      node[lbl] {normalize, $\to_{\mathbf R}^{8}$\\};
    \draw[flow] (l3.south) -- (l3.south |- lg.north)
      node[lbl] {normalize, $\to_{\mathbf R}^{9}$};
    \draw[flow] (qg.south) -- (qg.south |- idx.north) node[lbl] {color refinement,\\signature};
    \draw[flow] (lg.south) -- (lg.south |- idx.north) node[lbl] {color refinement,\\signature};
    \draw[flow] (idx) -- (iso);
    \draw[flow] (iso) -- (cert) node[lbl] {identified pair};
    \draw[flow] (cert) -- (dup) node[lbl] {accepted};
  \end{tikzpicture}%
  \end{lrbox}%
  \resizebox{\textwidth}{!}{\usebox{\pipelinebox}}
  \caption{Duplicate detection over a library $\mathcal L$ on the
    running example.  Every theorem has its reducible definitions
    unfolded (here only $m\mid n$) and is put into weak head normal
    form by Lean, embedded into \tm, normalized by \R (the
    individual rewrite steps are collapsed here), and inserted into an index keyed by the
    color-refinement signature of the DAG of its normal form.
    The fills of the nodes in $N_{T_1}$ and $N_{T_2}$ are the color
    classes after refinement; equal fills are equal color values, and
    the equation node carries no color.
    Only theorems in the same bucket are
    compared, by searching for a graph isomorphism modulo binder
    permutation.  $T_1$ and $T_2$ share a bucket and are isomorphic.
    For the identified pair a Lean proof of $T_1 \leftrightarrow T_2$ is
    built from the two rewrite derivations and the isomorphism and is
    checked by the Lean kernel; only then is the pair reported as a
    duplicate.}
  \label{fig:pipeline}
\end{figure}

%% file: parts/tables/tab_run.tex
\begin{table}[htbp]
  \centering
  \caption{One run of the tool over Lean core, Batteries and a pinned Mathlib.
    Every number is computed by the analysis scripts of the run from its outputs and reproducible with the provided artifact.
  Definitional equality is by the verdict of \texttt{Meta.isDefEq} at default transparency on the representative and the member after positional renaming of universe parameters.}
  \label{tab:run-details}
  \small
  \begin{tabular}{@{}lr@{}}
    \toprule
    theorem statements extracted (non-private) & 286{,}264 \\
    \quad encoded and normalized within the size budget & 286{,}224 \\
    \quad over budget, never compared & 40 \\
    \midrule
    duplicate classes (at least two members) & 7{,}525 \\
    \quad members in these classes & 15{,}834 \\
    pairs (representative, member), one kernel-checked certificate each & 8{,}309 \\
    \quad member definitionally equal to the representative & 6{,}293 (76\,\%) \\
    \quad member not definitionally equal to the representative & 2{,}016 \\
    \quad head pairs (representative, head of a definitional-equality group) & 1{,}912 \\
    \midrule
    pairs where both sides carry a proof of their own & 7{,}612 \\
    \quad and neither side is generated from the other & 3{,}003 \\
    \quad and neither proof references the other & 2{,}051 \\
    \quad and the sides are not joined through other members & 1{,}896 \\
    \quad and neither side is deprecated (independent pairs) & 1{,}742 \\
    \midrule
    redundant proofs (independent proofs $-$ 1 per class) & 1{,}747 in 1{,}687 classes \\
    \quad for definitionally equal statements / equal only through \iso & 965 / 782 \\
    \quad not labeled as deliberate API twins & 873 in 844 classes \\
    \quad\quad definitionally equal / equal only up to isomorphisms & 586 / 287 \\
    \quad additional findings with a deprecated side & 109 \\
    \quad relevant findings including deprecated ones & 1{,}856 \\
    \quad spanning packages & 192 in 162 classes \\
    \quad with no pair found by Mathlib's duplicate-declarations linter & 1{,}226 in 1{,}184 classes \\
    \midrule
    independent pairs found / excluded / missed by Mathlib's linter & 481 / 241 / 1{,}020 \\
    \bottomrule
  \end{tabular}
\end{table}

%% file: parts/listings/finding_list_zipidx.tex
\begin{Verbatim}[commandchars=\\\{\}]
\PYG{c+c1}{\PYGZhy{}\PYGZhy{} Init/Data/List/Nat/Range.lean}
\PYG{k+kn}{theorem}\PYG{+w}{ }\PYG{n}{map\PYGZus{}zipIdx}\PYG{+w}{ }\PYG{o}{\PYGZob{}}\PYG{n}{f}\PYG{+w}{ }\PYG{o}{:}\PYG{+w}{ }\PYG{n}{α}\PYG{+w}{ }\PYG{n+nb+bp}{→}\PYG{+w}{ }\PYG{n}{β}\PYG{o}{\PYGZcb{}}\PYG{+w}{ }\PYG{o}{\PYGZob{}}\PYG{n}{l}\PYG{+w}{ }\PYG{o}{:}\PYG{+w}{ }\PYG{n}{List}\PYG{+w}{ }\PYG{n}{α}\PYG{o}{\PYGZcb{}}\PYG{+w}{ }\PYG{o}{\PYGZob{}}\PYG{n}{k}\PYG{+w}{ }\PYG{o}{:}\PYG{+w}{ }\PYG{n}{Nat}\PYG{o}{\PYGZcb{}}\PYG{+w}{ }\PYG{o}{:}
\PYG{+w}{    }\PYG{n}{map}\PYG{+w}{ }\PYG{o}{(}\PYG{n}{Prod}\PYG{n+nb+bp}{.}\PYG{n}{map}\PYG{+w}{ }\PYG{n}{f}\PYG{+w}{ }\PYG{n}{id}\PYG{o}{)}\PYG{+w}{ }\PYG{o}{(}\PYG{n}{zipIdx}\PYG{+w}{ }\PYG{n}{l}\PYG{+w}{ }\PYG{n}{k}\PYG{o}{)}\PYG{+w}{ }\PYG{n+nb+bp}{=}\PYG{+w}{ }\PYG{n}{zipIdx}\PYG{+w}{ }\PYG{o}{(}\PYG{n}{l}\PYG{n+nb+bp}{.}\PYG{n}{map}\PYG{+w}{ }\PYG{n}{f}\PYG{o}{)}\PYG{+w}{ }\PYG{n}{k}\PYG{+w}{ }\PYG{o}{:=}\PYG{+w}{ }\PYG{k}{by}
\PYG{+w}{  }\PYG{n}{induction}\PYG{+w}{ }\PYG{n}{l}\PYG{+w}{ }\PYG{n}{generalizing}\PYG{+w}{ }\PYG{n}{k}\PYG{+w}{ }\PYG{n+nb+bp}{\PYGZlt{};\PYGZgt{}}\PYG{+w}{ }\PYG{n}{simp\PYGZus{}all}

\PYG{c+c1}{\PYGZhy{}\PYGZhy{} Init/Data/List/Nat/Range.lean, 31 lines further down}
\PYG{k+kn}{theorem}\PYG{+w}{ }\PYG{n}{zipIdx\PYGZus{}map}\PYG{+w}{ }\PYG{o}{\PYGZob{}}\PYG{n}{l}\PYG{+w}{ }\PYG{o}{:}\PYG{+w}{ }\PYG{n}{List}\PYG{+w}{ }\PYG{n}{α}\PYG{o}{\PYGZcb{}}\PYG{+w}{ }\PYG{o}{\PYGZob{}}\PYG{n}{k}\PYG{+w}{ }\PYG{o}{:}\PYG{+w}{ }\PYG{n}{Nat}\PYG{o}{\PYGZcb{}}\PYG{+w}{ }\PYG{o}{\PYGZob{}}\PYG{n}{f}\PYG{+w}{ }\PYG{o}{:}\PYG{+w}{ }\PYG{n}{α}\PYG{+w}{ }\PYG{n+nb+bp}{→}\PYG{+w}{ }\PYG{n}{β}\PYG{o}{\PYGZcb{}}\PYG{+w}{ }\PYG{o}{:}
\PYG{+w}{    }\PYG{n}{zipIdx}\PYG{+w}{ }\PYG{o}{(}\PYG{n}{l}\PYG{n+nb+bp}{.}\PYG{n}{map}\PYG{+w}{ }\PYG{n}{f}\PYG{o}{)}\PYG{+w}{ }\PYG{n}{k}\PYG{+w}{ }\PYG{n+nb+bp}{=}\PYG{+w}{ }\PYG{o}{(}\PYG{n}{zipIdx}\PYG{+w}{ }\PYG{n}{l}\PYG{+w}{ }\PYG{n}{k}\PYG{o}{)}\PYG{n+nb+bp}{.}\PYG{n}{map}\PYG{+w}{ }\PYG{o}{(}\PYG{n}{Prod}\PYG{n+nb+bp}{.}\PYG{n}{map}\PYG{+w}{ }\PYG{n}{f}\PYG{+w}{ }\PYG{n}{id}\PYG{o}{)}\PYG{+w}{ }\PYG{o}{:=}\PYG{+w}{ }\PYG{k}{by}
\PYG{+w}{  }\PYG{n}{induction}\PYG{+w}{ }\PYG{n}{l}\PYG{+w}{ }\PYG{k}{with}
\PYG{+w}{  }\PYG{n+nb+bp}{|}\PYG{+w}{ }\PYG{n}{nil}\PYG{+w}{ }\PYG{n+nb+bp}{=\PYGZgt{}}\PYG{+w}{ }\PYG{n}{rfl}
\PYG{+w}{  }\PYG{n+nb+bp}{|}\PYG{+w}{ }\PYG{n}{cons}\PYG{+w}{ }\PYG{n}{hd}\PYG{+w}{ }\PYG{n}{tl}\PYG{+w}{ }\PYG{n}{IH}\PYG{+w}{ }\PYG{n+nb+bp}{=\PYGZgt{}}
\PYG{+w}{    }\PYG{n}{rw}\PYG{+w}{ }\PYG{o}{[}\PYG{n}{map\PYGZus{}cons}\PYG{o}{,}\PYG{+w}{ }\PYG{n}{zipIdx\PYGZus{}cons\PYGZsq{}}\PYG{o}{,}\PYG{+w}{ }\PYG{n}{zipIdx\PYGZus{}cons\PYGZsq{}}\PYG{o}{,}\PYG{+w}{ }\PYG{n}{map\PYGZus{}cons}\PYG{o}{,}\PYG{+w}{ }\PYG{n}{map\PYGZus{}map}\PYG{o}{,}\PYG{+w}{ }\PYG{n}{IH}\PYG{o}{,}
\PYG{+w}{      }\PYG{n}{map\PYGZus{}map}\PYG{o}{]}
\PYG{+w}{    }\PYG{n}{rfl}
\end{Verbatim}

%% file: parts/listings/finding_tensorproduct.tex
\begin{Verbatim}[commandchars=\\\{\}]
\PYG{c+c1}{\PYGZhy{}\PYGZhy{} Mathlib/Algebra/Colimit/TensorProduct.lean}
\PYG{k+kn}{theorem}\PYG{+w}{ }\PYG{n}{Submodule}\PYG{n+nb+bp}{.}\PYG{n}{FG}\PYG{n+nb+bp}{.}\PYG{n}{exists\PYGZus{}rTensor\PYGZus{}fg\PYGZus{}inclusion\PYGZus{}eq}
\PYG{+w}{    }\PYG{o}{\PYGZob{}}\PYG{n}{N}\PYG{+w}{ }\PYG{o}{:}\PYG{+w}{ }\PYG{n}{Submodule}\PYG{+w}{ }\PYG{n}{R}\PYG{+w}{ }\PYG{n}{P}\PYG{o}{\PYGZcb{}}\PYG{+w}{ }\PYG{o}{(}\PYG{n}{hN}\PYG{+w}{ }\PYG{o}{:}\PYG{+w}{ }\PYG{n}{N}\PYG{n+nb+bp}{.}\PYG{n}{FG}\PYG{o}{)}\PYG{+w}{ }\PYG{o}{\PYGZob{}}\PYG{n}{x}\PYG{+w}{ }\PYG{n}{y}\PYG{+w}{ }\PYG{o}{:}\PYG{+w}{ }\PYG{n}{N}\PYG{+w}{ }\PYG{n+nb+bp}{⊗}\PYG{o}{[}\PYG{n}{R}\PYG{o}{]}\PYG{+w}{ }\PYG{n}{M}\PYG{o}{\PYGZcb{}}
\PYG{+w}{    }\PYG{o}{(}\PYG{n}{eq}\PYG{+w}{ }\PYG{o}{:}\PYG{+w}{ }\PYG{n}{N}\PYG{n+nb+bp}{.}\PYG{n}{subtype}\PYG{n+nb+bp}{.}\PYG{n}{rTensor}\PYG{+w}{ }\PYG{n}{M}\PYG{+w}{ }\PYG{n}{x}\PYG{+w}{ }\PYG{n+nb+bp}{=}\PYG{+w}{ }\PYG{n}{N}\PYG{n+nb+bp}{.}\PYG{n}{subtype}\PYG{n+nb+bp}{.}\PYG{n}{rTensor}\PYG{+w}{ }\PYG{n}{M}\PYG{+w}{ }\PYG{n}{y}\PYG{o}{)}\PYG{+w}{ }\PYG{o}{:}
\PYG{+w}{    }\PYG{n+nb+bp}{∃}\PYG{+w}{ }\PYG{n}{N\PYGZsq{}}\PYG{o}{,}\PYG{+w}{ }\PYG{n}{N\PYGZsq{}}\PYG{n+nb+bp}{.}\PYG{n}{FG}\PYG{+w}{ }\PYG{n+nb+bp}{∧}\PYG{+w}{ }\PYG{n+nb+bp}{∃}\PYG{+w}{ }\PYG{n}{h}\PYG{+w}{ }\PYG{o}{:}\PYG{+w}{ }\PYG{n}{N}\PYG{+w}{ }\PYG{n+nb+bp}{≤}\PYG{+w}{ }\PYG{n}{N\PYGZsq{}}\PYG{o}{,}
\PYG{+w}{      }\PYG{o}{(}\PYG{n}{N}\PYG{n+nb+bp}{.}\PYG{n}{inclusion}\PYG{+w}{ }\PYG{n}{h}\PYG{o}{)}\PYG{n+nb+bp}{.}\PYG{n}{rTensor}\PYG{+w}{ }\PYG{n}{M}\PYG{+w}{ }\PYG{n}{x}\PYG{+w}{ }\PYG{n+nb+bp}{=}
\PYG{+w}{        }\PYG{o}{(}\PYG{n}{N}\PYG{n+nb+bp}{.}\PYG{n}{inclusion}\PYG{+w}{ }\PYG{n}{h}\PYG{o}{)}\PYG{n+nb+bp}{.}\PYG{n}{rTensor}\PYG{+w}{ }\PYG{n}{M}\PYG{+w}{ }\PYG{n}{y}\PYG{+w}{ }\PYG{o}{:=}\PYG{+w}{ }\PYG{k}{by}\PYG{+w}{ }\PYG{n+nb+bp}{...}

\PYG{c+c1}{\PYGZhy{}\PYGZhy{} Mathlib/RingTheory/TensorProduct/DirectLimitFG.lean}
\PYG{k+kn}{theorem}\PYG{+w}{ }\PYG{n}{TensorProduct}\PYG{n+nb+bp}{.}\PYG{n}{eq\PYGZus{}of\PYGZus{}fg\PYGZus{}of\PYGZus{}subtype\PYGZus{}eq}\PYG{+w}{ }\PYG{o}{\PYGZob{}}\PYG{n}{t\PYGZsq{}}\PYG{+w}{ }\PYG{o}{:}\PYG{+w}{ }\PYG{n}{P}\PYG{+w}{ }\PYG{n+nb+bp}{⊗}\PYG{o}{[}\PYG{n}{R}\PYG{o}{]}\PYG{+w}{ }\PYG{n}{N}\PYG{o}{\PYGZcb{}}
\PYG{+w}{    }\PYG{o}{(}\PYG{n}{h}\PYG{+w}{ }\PYG{o}{:}\PYG{+w}{ }\PYG{n}{rTensor}\PYG{+w}{ }\PYG{n}{N}\PYG{+w}{ }\PYG{n}{P}\PYG{n+nb+bp}{.}\PYG{n}{subtype}\PYG{+w}{ }\PYG{n}{t}\PYG{+w}{ }\PYG{n+nb+bp}{=}\PYG{+w}{ }\PYG{n}{rTensor}\PYG{+w}{ }\PYG{n}{N}\PYG{+w}{ }\PYG{n}{P}\PYG{n+nb+bp}{.}\PYG{n}{subtype}\PYG{+w}{ }\PYG{n}{t\PYGZsq{}}\PYG{o}{)}\PYG{+w}{ }\PYG{o}{:}
\PYG{+w}{    }\PYG{n+nb+bp}{∃}\PYG{+w}{ }\PYG{o}{(}\PYG{n}{Q}\PYG{+w}{ }\PYG{o}{:}\PYG{+w}{ }\PYG{n}{Submodule}\PYG{+w}{ }\PYG{n}{R}\PYG{+w}{ }\PYG{n}{M}\PYG{o}{)}\PYG{+w}{ }\PYG{o}{(}\PYG{n}{hPQ}\PYG{+w}{ }\PYG{o}{:}\PYG{+w}{ }\PYG{n}{P}\PYG{+w}{ }\PYG{n+nb+bp}{≤}\PYG{+w}{ }\PYG{n}{Q}\PYG{o}{),}\PYG{+w}{ }\PYG{n}{Q}\PYG{n+nb+bp}{.}\PYG{n}{FG}\PYG{+w}{ }\PYG{n+nb+bp}{∧}
\PYG{+w}{      }\PYG{n}{rTensor}\PYG{+w}{ }\PYG{n}{N}\PYG{+w}{ }\PYG{o}{(}\PYG{n}{inclusion}\PYG{+w}{ }\PYG{n}{hPQ}\PYG{o}{)}\PYG{+w}{ }\PYG{n}{t}\PYG{+w}{ }\PYG{n+nb+bp}{=}
\PYG{+w}{        }\PYG{n}{rTensor}\PYG{+w}{ }\PYG{n}{N}\PYG{+w}{ }\PYG{o}{(}\PYG{n}{inclusion}\PYG{+w}{ }\PYG{n}{hPQ}\PYG{o}{)}\PYG{+w}{ }\PYG{n}{t\PYGZsq{}}\PYG{+w}{ }\PYG{o}{:=}\PYG{+w}{ }\PYG{k}{by}\PYG{+w}{ }\PYG{n+nb+bp}{...}
\end{Verbatim}